\documentclass{article}

\usepackage{arxiv}

\usepackage[english]{babel}

\usepackage[utf8]{inputenc} 
\usepackage[T1]{fontenc}    
\usepackage{amsmath}
\usepackage{graphicx}
  \DeclareGraphicsExtensions{.pdf,.png}
\usepackage{pgfplots}
\pgfplotsset{compat=1.18}
\usepackage{authblk}

\usepackage{amssymb}  
\usepackage{pifont} 

\usepackage[colorlinks=true, allcolors=blue]{hyperref}
\usepackage{apacite}
\usepackage[most]{tcolorbox}
\usetikzlibrary{positioning, shapes}
\def\BibTeX{{\rm B\kern-.05em{\sc i\kern-.025em b}\kern-.08em
    T\kern-.1667em\lower.7ex\hbox{E}\kern-.125emX}}

\makeatletter
\renewcommand\paragraph{\@startsection{paragraph}{4}{\z@}%
  {3.25ex \@plus1ex \@minus.2ex}%
  {-1em}%
  {\normalfont\normalsize\itshape}}
\makeatother

\begin{document}

\title{Opacity as a Feature, Not a Flaw: The LoBOX Governance Ethic for Role-Sensitive Explainability and Institutional Trust in AI}

\newcommand{\shorttitle}{The LoBOX Governance Ethic Framework} 

\author{Francisco Herrera$^1$$^2$, Reyes Calderón$^3$}
\affil{$^1$Department of Computer Science and Artificial Intelligence, Andalusian Institute of Data Science and Computational Intelligence (DaSCI), University of Granada, Spain. \\ \texttt{herrera@decsai.ugr.es} }
\affil{$^2$ ADIA Lab, Abu Dhabi,United Arab Emirates}

\affil{$^3$Department of Management and IIT, University Pontificia de Comillas, Madrid, Spain. \\ \texttt{mrcalderon@comillas.edu} }

\maketitle

\begin{abstract}

This paper introduces LoBOX (Lack of Belief: Opacity \& eXplainability) governance ethic structured framework for managing artificial intelligence (AI) opacity when full transparency is infeasible. Rather than treating opacity as a design flaw, LoBOX defines it as a condition that can be ethically governed through role-calibrated explanation and institutional accountability. The framework comprises a three-stage pathway: reduce accidental opacity, bound irreducible opacity, and delegate trust through structured oversight. Integrating the RED/BLUE XAI model for stakeholder-sensitive explanation and aligned with emerging legal instruments such as the EU AI Act, LoBOX offers a scalable and context-aware alternative to transparency-centric approaches. Reframe trust not as a function of complete system explainability, but as an outcome of institutional credibility, structured justification, and stakeholder-responsive accountability.  A governance loop cycles back to ensure that LoBOX remains responsive to evolving technological contexts and stakeholder expectations, to ensure the complete opacity governance.  We move from transparency ideals to ethical governance, emphasizing that trustworthiness in AI must be institutionally grounded and contextually justified. We also discuss how cultural or institutional trust varies in different contexts. This theoretical framework positions opacity not as a flaw but as a feature that must be actively governed to ensure responsible AI systems.
\end{abstract}



\begin{keywords} 
~
Explainable Artificial Intelligence (XAI), Algorithmic opacity, AI governance, Institutional trust, Role-sensitive explainability, Accountability
\end{keywords}

\newpage

\section*{Highlights}
\begin{itemize}
\item Introduces LoBOX (Lack of Belief: Opacity \& eXplainability) governance ethic framework to  manage opacity in AI systems
\item 	Challenges the dominance of transparency-based paradigms, reframing opacity as a condition to be ethically governed—not eliminated.
\item 	Proposes a three-stage governance pathway: (1) reduce accidental opacity, (2) bound irreducible opacity, and (3) delegate trust through institutional oversight, including a governance loop back. 
\item Integrates the RED/BLUE XAI model to align explanation strategies with stakeholder roles and through role-sensitive explainability.
\item 	Emphasizes institutional scaffolding and stakeholder-contingent trust as alternatives to universal explainability.
\end{itemize}
\newpage

\section{Introduction}

The metaphor of the “black box” has become the dominant topic in public and academic discourse on artificial intelligence (AI) \cite{Castelvecchi2016}, reflecting the discomfort with systems whose internal workings resist scrutiny and accountability. Transparency is often framed as the antidote: if we could open the box and see how decisions are made, AI could be rendered fair, safe, and trustworthy. Yet, this perspective, however well-intentioned, rests on a flawed assumption: that transparency is always feasible and ethically sufficient.

Explainable AI (XAI) has emerged as the central transparence field tasked with making opaque systems intelligible and accountable. Initially focused on technical transparency and model interpretability, XAI has evolved into a governance-oriented discipline that bridges the cognitive complexity of AI with the ethical and legal responsibilities of its users. As Arrieta et al. (2020) \cite{arrieta2020explainable} emphasize, XAI is essential to foster trust and responsible Ai, based on two central concepts: understanding and audience. However, as Herrera (2025) \cite{herrera2025reflections} notes, explainability must go beyond surface-level transparency and embrace its relational and contextual dimensions, including support for human-AI collaboration.

Within this evolving landscape, Role-Sensitive Explainability has emerged as a foundational principle. It recognizes that not all users need the same kind of explanation: explanations must be tailored to the cognitive needs, stakes, and institutional roles of diverse stakeholders, from developers and regulators to experts and lay users. This shift repositions XAI as a socio-technical bridge rather than a generic toolkit, one that connects AI systems to their contexts of use through adaptive, audience-calibrated strategies. The RED / BLUE XAI approach -Research, Explore, Debug, and responsiBle, Legal, trUst, Ethics - XAI \cite{biecek2024explain} reflects this paradigm. RED XAI supports developers and auditors through technical fidelity and diagnostic transparency, while BLUE XAI provides lay users and decision makers with accessible, ethically grounded justifications \cite{biecek2024explain,herrera2025reflections}.

Governance, therefore, must tailor explanations to the epistemic role, risk level, and decision-making authority \cite{khalili2024against,tielman2024explainable}. Moreover, in high-risk or opaque contexts where even structured explanations can fail, the foundation of public trust must shift to institutions that are credible, transparent in their procedures, and responsive to public disputes \cite{freiman2025opacity,de2024black}. This reflects the larger movement from epistemic idealism toward ethical justification and procedural legitimacy, where trust is earned not by making everything visible, but by making governance accountable and explanations meaningful \cite{van2023critique,mollering2001nature,flores1998creating}.

A central hypothesis of governance beyond transparency is that the pursuit of total transparency, the belief that all internal mechanisms of a system should be fully visible and universally understandable, is both unrealistic and ethically insufficient. In complex sociotechnical systems, particularly those powered by deep learning, internal operations may be technically accessible but remain cognitively opaque even to domain experts \cite{boge2022two,lipton2018mythos}. Full transparency can mislead non-experts by overwhelming them with uninterpretable data, expose sensitive intellectual property, or create an illusion of fairness while concealing deeper systemic risks \cite{ananny2018seeing,hutson2021opacity,selbst2019fairness}. As a result, transparency often fails to deliver real accountability or informed trust. Instead, governance must aim for structured intelligibility: a deliberate, stakeholder-specific effort to make systems understandable enough to enable scrutiny, contestation, and ethically meaningful decision-making \cite{fraser2022ai}.

This paper challenges that assumption by offering a new perspective: opacity should be treated not as a failure, but as an ethically governable challenge. We argue that opacity is not inherently unethical nor is transparency inherently virtuous. Instead, opacity should be reframed as a governable condition, if approached with nuance and institutional care, and it can be rendered ethically justifiable.  

Therefore, this paper advances a reflection from transparency as default to partical opacity as a governance possibility, embedding ethics not in visibility, but in structured justification and stakeholder-calibrated oversight. We move from transparency ideals to ethical governance, emphasizing that trustworthiness in AI must be institutionally grounded and contextually justified. In doing so, we propose a shift from transparency to governable opacity, where the explanation is tailored to roles, risks, and social responsibility, rather than being treated as a universal remedy. Our perspective discussion and proposal align with interdisciplinary critiques of transparency as an oversimplified value \cite{lipton2018mythos,ananny2018seeing,selbst2019fairness}, and resonate with governance models that prioritize contestability and oversight \cite{freiman2025opacity,raji2020closing,fraser2022ai}.

Following the previous hypothesis, this paper introduces the LoBOX governance ethic framework - Lack of Belief: Opacity \& eXplainability — as a framework to manage opacity through structured justification, role-sensitive explainability, and institutional accountability. It builds on the recognition that opacity in AI is often inevitable, especially in systems that rely on high-dimensional nonlinear models such as deep learning. The lack of belief in AI systems, by users, regulators, or the public, often stems from opacity and insufficient explainability.

LoBOX reframes opacity not as a flaw to be eliminated but as a condition to be ethically governed, particularly in high-risk  domains where explanation is inherently partial.  The critical question is not whether we can see inside the machine, but whether affected stakeholders can challenge outcomes and trust the institutions that govern them. In this light, explainability is not merely a technical aspiration but a normative commitment, one that aligns explanation with relational, institutional, and moral accountability. This framework moves us beyond the unattainable ideal of full transparency and toward pragmatic opacity ethics grounded in stakeholder-calibrated intelligibility, contextual trust, and institutional legitimacy.  A governance loop cycles back ensure that LoBOX remains responsive to evolving technological contexts and stakeholder expectations, to ensure the complete opacity governance. 

The remainder of this paper is structured as follows. Section \ref{sec:opacity} develops the theoretical foundation for LoBOX, introducing opacity as a frontier of ethical design. It distinguishes between accidental and per se opacity and frames the explanation as a role-sensitive governance tool. Section \ref{sec:LoBOX} presents the LoBOX governance ethic in detail, highlighting its three-stage pathway: reducing accidental opacity, bounding irreducible opacity, and delegating trust through institutional oversight, together with a governance loop back. It also highlights LoBOX’s integration with the RED/BLUE XAI model and its alignment with emerging legal standards, particularly the EU AI Act. Section \ref{sec:App} presents three practical scenarios using LoBOX. Section \ref{sec:society} discusses how cultural or institutional trust varies between contexts.  Section \ref{sec:con} presents the final conclusions and reflections. 

\section{Opacity as a Frontier of  Ethical Design}
\label{sec:opacity}

Understanding opacity as an ethical challenge rather than a technical flaw requires a change in the way we conceptualize explanation, intelligibility, and institutional responsibility. This section develops that shift across four dimensions. Subsection \ref{sec:opacity-1} introduces the concept of role-relative intelligibility, arguing that effective explainability must be calibrated to the epistemic roles of stakeholders. Subsection \ref{sec:opacity-2} distinguishes between accidental and per se opacity, showing how each type demands distinct governance responses. Section \ref{sec:opacity-3} challenges the ideal of exhaustive transparency from an epistemic realist perspective, proposing an explanation as a contextual tool for public accountability rather than universal understanding. Finally, Subsection \ref{sec:opacity-4} situates opacity within a broader ethic of institutional trust, reframing the "leap of faith" not as a psychological concession, but as a governance obligation that the LoBOX governance ethics framework helps to structure.

\subsection{From Transparency Idealism to Role-Relative Intelligibility}
\label{sec:opacity-1}

Opacity in AI systems is not merely a technical property: it is relational, shaped by who observes, in what context, and with what expectations. A system may appear opaque to a layperson, yet interpretable to an engineer or auditable by a regulator. Effective governance must therefore account for the cognitive distance, the epistemic gap between what a system processes and what a stakeholder can reasonably understand \cite{herrera2025reflections}.

Instead of pursuing universal transparency, explainability should aim at audience-relative intelligibility. Developers may need granular output for debugging, while end-users benefit more from high-level justifications that enable contestability. The explanation must be calibrated to the role and context, not derived from a one-size-fits-all model.
As Lipton (2018) \cite{lipton2018mythos} and Ananny and Crawford (2018) \cite{ananny2018seeing} warn, 'interpretability' often serves as a rhetorical placeholder, offering the illusion of understanding while concealing power asymmetries. Hutson (2021) \cite{hutson2021opacity} -discussing on AI systems opacity- calls this explainability theater: explanation interfaces that appear fair while masking bias. These critiques underscore the limits of superficial transparency and the need to shift toward moral accountability.

This perspective reframes the explanation not as an epistemic endpoint but as a strategic scaffolding of trust. Governance maturity models such as PAG-XAI \cite{herrera2025reflections} embrace this change: prioritizing actionable, audience-relevant intelligibility over idealized transparency. Such frameworks prepare the ground for the LoBOX governance ethic framework, which enables ethical deployment even when full comprehension is impossible.

\paragraph{\textbf{Illustration: Welfare algorithms and accidental opacity}} A national welfare agency uses an algorithm to assess benefit eligibility. Applicants often receive automated rejections with no explanation, sparking backlash and claims of unfairness. An independent audit reveals accidental opacity: unclear language, missing documentation, and no appeal process. The agency responds by redesigning the interface: specific denial reasons, rules references, and accessible appeal pathways. Although the system remains complex, these RED XAI interventions make decisions intelligible and contestable, providing transparency where it matters most.

\subsection{Typologies of Opacity: Accidental and Per Se}
\label{sec:opacity-2}

Boge (2022) \cite{boge2022two} identify two corresponding dimensions: opacity of training and opacity of representation.

\begin{itemize}
    \item The opacity of training, linked to the complexity of model development, data pre-processing, and hyperparameter optimization, is closely related to what we describe as accidental opacity. 
    
     \item In contrast, opacity of representation, which concerns the inscrutability of internal model states such as weights and activations, reflects the core of per se opacity. Crucially, Boge argues that many deep learning systems are instrumental, they produce valuable outputs without furnishing intelligible internal reasoning, highlighting the limits of explanation as a route to understanding. 

\end{itemize}

This reinforces our view that ethical governance cannot rely on epistemic transparency alone; rather, it must scaffold trust through institutional structures and morally calibrated oversight (\cite{boge2022two}. pp. 47–54).

Calls for transparency often overlook the different sources of opacity. We propose a conceptual distinction between two principal forms of opacity connecting with the LoBOX governance ethic framework: 

\begin{itemize}
    \item \textbf{Accidental opacity}  arises from avoidable conditions such as poor user interface design, inadequate documentation, or institutional secrecy. It is typically the result of inadequate communication or misaligned organizational incentives and can often be remedied through improved practices in system design, documentation, or transparency.
    
\item \textbf{Per se opacity}, on the contrary, is intrinsic to the complexity of certain AI models, particularly deep learning systems, whose internal logic may be formally accessible, but remains cognitively opaque even to expert users. Here, the challenge is not poor design, but the limits of human interpretability in the face of statistical abstraction and high-dimensional computation.
\end{itemize}

These two forms of opacity require different responses. Where accidental opacity can and should be reduced through targeted interventions, per se opacity requires ethical management rather than elimination. This management includes interpretability tools, layered or role-specific explanations, and formal accountability structures (governance ethic structures) that render the system justifiable even if not fully intelligible.

\paragraph{\textbf{Illustration: Radiology AI and the ethics of per-sequence opacity}} A deep learning model helps radiologists detect lung abnormalities from chest X-rays. The internal logic of the model is not interpretable, even for system developers - typifying per se opacity. Although engineers use proxy interpretability tools to monitor model drift (RED/BLUE XAI), the system is governed by institutional protocols that restrict its autonomous use and document each diagnostic decision. Patients, meanwhile, are given high-level explanations in nontechnical language and retain the right to consult a human expert. In this setting, opacity is not eliminated, but ethically scaffolded: through layered explanation, stakeholder-sensitive design, and institutional trust. This exemplifies the LoBOX governance ethic framework, where transparency is distributed and accountability is preserved through governance.

\subsection{The Case Against Ideal Transparency: An Epistemic Realist View}
\label{sec:opacity-3}

Bodria et al.(2023) \cite{bodria2023benchmarking} provide a comprehensive taxonomy of explanation methods for black-box AI models, categorizing them based on the type of explanation they produce and the data formats they handle. This taxonomy aims to guide researchers and practitioners in selecting appropriate explanation methods tailored to specific use cases.

Prevailing calls for transparency in AI frequently rest on the assumption that explanations should be exhaustive, intuitively comprehensible, and universally accessible. This assumption, here referred to as epistemic idealism, fails to acknowledge both the structural complexity of contemporary AI systems and the cognitive constraints of their diverse users. Such an approach risks substituting genuine understanding with superficial clarity, relying on simplified narratives or visualizations that obscure rather than elucidate system behavior. In practice, explanation is inherently partial and context-dependent, shaped by interpretive capacities, institutional roles, and epistemic stakes of individuals and groups who engage with the system.

In contrast, epistemic realism accepts that explainability is not an inherent property of a system, but an outcome of social, institutional, and moral context. What matters is not whether everyone can understand everything, but whether each stakeholder can understand enough to make informed, justifiable, and accountable judgments. From this perspective, explanation is not a window into machine logic; it is a tool to make publicly acceptable decisions.

When transparency alone is insufficient to build trust, governance must instead depend on structured, role-sensitive forms of justification. Selbst et al. (2019) \cite{selbst2019fairness} highlight how ethical failure often stems not from opacity but from abstraction, the detachment of AI systems from the contexts in which they operate. An effective explanation, they argue, must be grounded in these real-world contexts, not abstract ideals. The RED/Blue XAI approach addresses this directly by aligning explanation strategies with stakeholder-sensitive needs and system risks.

Freiman et al. (2025) \cite{freiman2025opacity} extend this argument by distinguishing between technical, institutional, and epistemic opacity, showing that trust cannot rely on transparency alone. Instead, it must be scaffolded through governance structures calibrated to the role and context of the stakeholder, a principle at the heart of the LoBOX governance ethic framework.

Some critics may argue that explanation is a moral right and that opacity inherently undermines autonomy. Although we acknowledge this concern, it often overlooks the limitations of explanation in high-complexity systems. Vaassen (2022) \cite{vaassen2022ai} clarifies this by showing that what matters is not full access to internal workings, but whether individuals can challenge and respond to decisions that affect them. Similarly, Khalili (2024) \cite{khalili2024against} argues that what allows moral reasoning is not mechanical transparency but qualitative understanding of the ability to relate AI behavior to real-world consequences. This is connected with the definition of XAI provided by Arrieta et al. (2020) \cite{arrieta2020explainable} with understanding and audience playing the fundamental role.

Together, these perspectives affirm our core claim. 

\begin{quote}
    "Truly honest opacity is not an ethical failure, it is an institutional achievement. When explanation is role-sensitive, morally relevant and embedded in oversight and recourse, opacity becomes not a barrier but a prerequisite for ethically viable AI governance."
\end{quote}

\subsection{Institutional Trust and the Leap of Faith}
\label{sec:opacity-4}

In democratic societies, trust in AI is grounded less in technical clarity than in institutional legitimacy. Just as citizens need not understand the biochemistry of vaccines to trust public health, users should not grasp the inner workings of neural networks to trust AI systems. Trust is sustained when systems are embedded in frameworks that ensure oversight, contestability, and morally relevant explanation.

This is especially crucial under per se opacity, where interpretability is inherently limited. In such cases, exhaustive transparency is neither feasible nor sufficient. Instead, mechanisms such as audits, ethics review boards, and third-party certification shift trust from system transparency to procedural legitimacy. Explanations serve not to completely open the black box, but to ensure intelligibility where it matters: clarity of the outcome, justification of the impact, and institutional accountability.

Legal systems provide instructive analogies. As Fraser et al. (2022) \cite{fraser2022ai} note, courts often rely on discovery, expert testimony, and parsimonious explanation - a clarity calibrated to the ethical or legal context. Similarly, de Andrade and Alves (2024) \cite{de2024black} demonstrate that targeted explanation, through tools like SHAP and LIME, can reveal bias and enable scrutiny, even in black-box systems. These cases confirm that opacity is not inherently unethical; what matters is whether it is governed.

This framework is further enriched by the critique of utilitarian trust offered by Van Rietschoten and Van Bommel (2023) \cite{van2023critique}, who argue that trust in modern institutions cannot be reduced to rational risk–benefit calculations.  It resonates with the critique of utilitarian trust advanced by the authors, who argue that in complex sociotechnical systems, stakeholders often cannot rationally assess risk or utility. Instead, they are placed in what the authors describe as a "leap of faith": a condition of trust without full understanding, a moment where trust cannot be grounded in knowledge and must instead be delegated to the integrity of the surrounding institutions of the system.

LoBOX reframes this leap not as a cognitive failure but as a governance opportunity: to construct institutionally scaffolded trust through role-sensitive explainability and layered oversight. Changes the burden of trust from individual comprehension to procedural and institutional credibility. In doing so, it positions opacity not as an ethical failure but as a condition to be responsibly managed. Within its third stage of governance, delegating trust through institutional oversight, trust is earned not through technical transparency but through processes that ensure accountability, contestability, and justifiability. In high-risk scenarios, this becomes the foundation for ethical legitimacy.

\section{Operationalizing Ethical Governance: The Three-Stage LoBOX Framework Pathway for Managing Opacity}
\label{sec:LoBOX}

Recognizing that not all opacity is equal and that different types of opacity require different responses, LoBOX structures governance around a scalable and role-sensitive pathway. This pathway enables AI developers, auditors, and regulators to intervene meaningfully at different points of opacity emergence, from design through deployment and oversight.

This section elaborates the core elements of the LoBOX governance ethic framework through five interconnected subsections. \ref{sec:LoBOX-1}  introduces the three-stage LoBOX framework, detailing its progression from reducing accidental opacity, to bounding irreducible opacity, and to delegating trust via institutional oversight, including a governance loop that cycles back. Subection \ref{sec:LoBOX-2} grounds the framework in philosophical and empirical critiques of utilitarian trust, demonstrating how LoBOX addresses the “leap of faith” as a governance obligation. \ref{sec:LoBOX-3} distills four key features that define the framework, including role-sensitive explainability, stakeholder-calibrated design, trustworthy opacity, and the governed leap of faith. \ref{sec:LoBOX-4} explains how LoBOX integrates the RED/BLUE XAI model to deliver layered intelligibility between stakeholder groups. Finally, \ref{sec:LoBOX-5} connects the framework to emerging legal requirements, most notably the EU AI Act.

Together, these subsections illustrate how LoBOX reframes the governance of opaque systems, not as a search for perfect transparency but as a commitment to structured justification, role-based accountability, and institutional legitimacy.

\subsection{LoBOX framework: Design, Deployment, Oversight and Governance Loop Cycles Back}
\label{sec:LoBOX-1}

In this section, we describe the three stages, together with a back-up of the governance loop.  The cycle ensures that LoBOX remains responsive to evolving technological contexts and stakeholder expectations to ensure complete opacity governance.

We propose the three-stage pathway for guiding the ethical deployment of opaque AI systems as:

\begin{itemize}
 
\item \textit{Stage 1 – Reduce Accidental Opacity}: Implement during design and prototyping by requiring user-centered interface evaluations, clearer model documentation, and transparency-by-design principles. It can also be developed along the training and validation, including RED XAI. This may include usability testing or pre-deployment review of decision rationales.

\item \textit{Stage 2 – Monitor Per Se Opacity}: Integrate interpretability checkpoints (e.g. SHAP, LIME, counterfactual analysis) during validation by experts and deployment. These checkpoints help ensure that performance audits, fairness assessments, and ethical boundary setting can proceed even when a full understanding of model internals is not feasible.

\item \textit{Stage 3 - Delegate Trust through Institutional Oversight}: Apply through external auditing cycles, red-teaming evaluations, inspection and safety AI, mechanisms for contestability and redress, and third-party ethics boards. In high-risk domains, these processes can align with legal requirements such as EU AI Act compliance or sector-specific AI certification standards.
  
\end{itemize}

To operationalize the LoBOX governance pathway, we synthesize its three stages into a tiered model that aligns explanation demands with the type of risk and opacity of the system. Table~\ref{tab:LoBOX-01} provides a structured overview of these stages, highlighting their focus, purpose, and example practices throughout the AI lifecycle.

\begin{table}[th!]
\centering
\caption{ LoBOX framework governance stages}
\label{tab:LoBOX-01}
\begin{tabular}{|p{2.5cm}|p{4cm}|p{5cm}|}
\hline

Stage & Description & Example Practices \\ \hline

Stage 1 – Reduce Accidental Opacity & Focuses on the design, prototyping phase and training to minimize avoidable opacity. & User-centered interface evaluations, clear documentation, transparency-by-design, usability testing. \\ \hline

Stage 2 – Monitor Per Se Opacity & Addresses irreducible opacity through interpretability tools during model validation and deployment. & Use of SHAP, LIME, counterfactual analysis; performance audits; ethical boundary-setting. \\ \hline

Stage 3 – Delegate Trust through Institutional Oversight & Shifts trust to institutional structures when interpretability is not possible. & External audits, red-teaming, inspection \& AI safety,  contestability mechanisms, alignment with EU AI Act. \\ \hline

\end{tabular}
\end{table}

In the following the development in these stages, paying attention to focus, strategies, and objective.

\section*{Stage 1 – Reduce Accidental Opacity}

\begin{itemize}

 \item[] Focus: Minimize the avoidable sources of opacity introduced by poor interface design, inadequate documentation, or unclear communication, issues that arise not from model complexity, but from insufficient attention to user understanding and system transparency.

  \item[]   Strategies:

\begin{itemize}
 
  \item Implement user-centered design to ensure interfaces support intuitive understanding.

  \item Develop clear, accessible documentation that describes model objectives, inputs, and limitations in non-technical language.

  \item Promote transparency-by-design, embedding clarity into the system architecture from the outset.

  \item Apply RED XAI techniques to provide internal stakeholders (e.g. developers, auditors) with technically detailed model explanations for debugging and compliance. Including data quality and data improvement, advancing towards data explainability to reduce accidental opacity.

\end{itemize}

 \item[] Objective: Prevent unnecessary confusion and distrust by ensuring that users, designers, and auditors can understand the system to the degree that is technically and cognitively feasible, including the prevention of fairness data and the discrimination actions if possible, etc.

\end{itemize}

\section*{Stage 2 – Bounding Irreducible Opacity}

\begin{itemize}

 \item[]
Focus: Address the unavoidable opacity inherent in complex AI models (e.g., deep learning) that cannot be fully interpreted, even by experts. This stage ensures that such models are ethically constrained and used only within justifiable boundaries.

 \item[]  Strategies:

\begin{itemize}
 
  \item 
Incorporate formal interpretability tools (e.g., SHAP, LIME, counterfactual explanations) to approximate how decisions are made.

\item Define explicit usage boundaries, specifying conditions under which the model is valid and safe to use, even if possible the data domain-analysis.

\item Maintain risk documentation and ethical constraints that acknowledge where explanation is limited and identify safeguards.

\item Use combined RED/Blue XAI to offer both technical insight and high-level moral explanations to internal and external stakeholders.

\end{itemize}

 \item[] Objective: Enable responsible deployment despite epistemic limits, by bounding the use of inherently opaque models within clear ethical and operational constraints, and ensuring justification through approximate intelligibility and governance tools.
\end{itemize}

\section*{Stage 3 – Delegate Trust through Institutional Oversight}

\begin{itemize}
 
  \item[] Focus: Reframe trust as a function of institutional legitimacy when full interpretability is impossible. Trust is not grounded in visibility, but in the credibility, transparency, and responsiveness of oversight structures.

 \item[]  Strategies:

\begin{itemize}
 
  \item
  Establish institutional audits and redress mechanisms to review model impacts and allow affected individuals to challenge decisions.

 \item  Install and activate inspection and AI safety models. 
 -
 \item Ensure human oversight pathways, empowering qualified professionals to monitor, intervene in, or override automated decisions.

 \item  BLUE XAI methods to deliver normatively meaningful, accessible explanations to stakeholders, allowing contestability and informed consent.
\end{itemize}

 \item[] Objective: Shift the foundation of public trust from system comprehension to procedural legitimacy. Where the explanation is partial or abstract, institutions must fill the gap with transparent governance, ethical justification, and enforceable accountability.
\end{itemize}

Based on the LoBOX tiered intervention logic, Table~\ref{tab:LoBOX-02} synthesizes the core attributes of each governance stage, summarizing their different focus areas, implementation strategies, and ethical objectives to guide practical implementation in varying AI risk contexts.

\begin{table}[th!]
\centering
\caption{ LoBOX governance stages summary: Focus, strategies and objective}
\label{tab:LoBOX-02}
\begin{tabular}{|p{2cm}|p{3.6cm}|p{3.4cm}|p{3.2cm}|}
\hline

Stage & Focus & Strategies & Objective \\ \hline

Stage 1 – Reduce Accidental Opacity & Minimize avoidable opacity from poor design, documentation, or communication; focus on user comprehension and system transparency. Paying attention to data quality & User-centered design; accessible documentation; transparency-by-design; RED XAI for technical stakeholders. & Prevent confusion and reduce reliance on blind trust through accessible system understanding. \\ \hline

Stage 2 – Bound Per Se (Irreducible) Opacity & Address unavoidable opacity in complex models; ensure such models are ethically constrained and justifiably deployed. & Interpretability tools (SHAP, LIME); define usage boundaries; risk documentation; combined RED/BLUE XAI. & Enable responsible use of opaque models via ethical boundaries and approximate intelligibility. \\ \hline

Stage 3 – Delegate Trust via Institutional Oversight & Reframe trust as institutional legitimacy; ensure oversight when full interpretability is not feasible. & Audits and redress mechanisms; human oversight pathways; BLUE XAI for public-facing intelligibility. & Shift trust from comprehension to procedural legitimacy through credible oversight and contestability. \\ \hline

\end{tabular}
\end{table}

To ensure that LoBOX remains responsive to evolving technological contexts and stakeholder expectations, we propose the integration of a governance feedback loop, a cyclical process of evaluation, adaptation, and calibration. Rather than treating LoBOX stages as static interventions, this loop emphasizes continuous monitoring of system behavior, stakeholder reactions, and institutional outcomes. Feedback from audits, user complaints, algorithmic failures, or changes in public trust can trigger adjustments in oversight mechanisms, explanation practices, or system deployment boundaries. This iterative governance model reinforces LoBOX’s adaptability by embedding a learning process within its structure, where trust is not only scaffolded, but maintained through transparent revision. In particular, in dynamic environments such as content moderation, predictive surveillance, or health diagnostics, the feedback loop ensures that ethical AI governance remains attuned to context, not just compliance.

When the feedback loop reengages an earlier stage of the LoBOX cycle, the nature of the intervention depends on the source and severity of the feedback. For example, if new evidence reveals that a previously 'bounded' system exhibits emergent biases or unintended harms, the framework must return to Stage 2 (Bounding) with expanded safeguards, revised deployment criteria, or enhanced explainability tools. Alternatively, if stakeholder trust erodes due to insufficient clarity or recourse.  The framework may return to Stage 1 (Reduction) to redesign user interfaces, communication protocols, or documentation, and quality data to reduce accidental opacity. In more systemic cases, such as regulatory noncompliance or widespread public backlash, the feedback loop may even require a temporary rollback of Stage 3 (Delegation), reasserting tighter institutional control or third-party audit mechanisms. This recursive capacity ensures that LoBOX is not merely reactive but anticipatory, capable of learning from system use in real time and recalibrating governance layers accordingly. By embedding iteration into the heart of its design, the LoBOX supports a living model of algorithmic accountability, responsive to risk, attentive to power asymmetries, and grounded in evolving public values.

\begin{figure*}[ht!]
    \centering
    \includegraphics[width=0.98\textwidth]{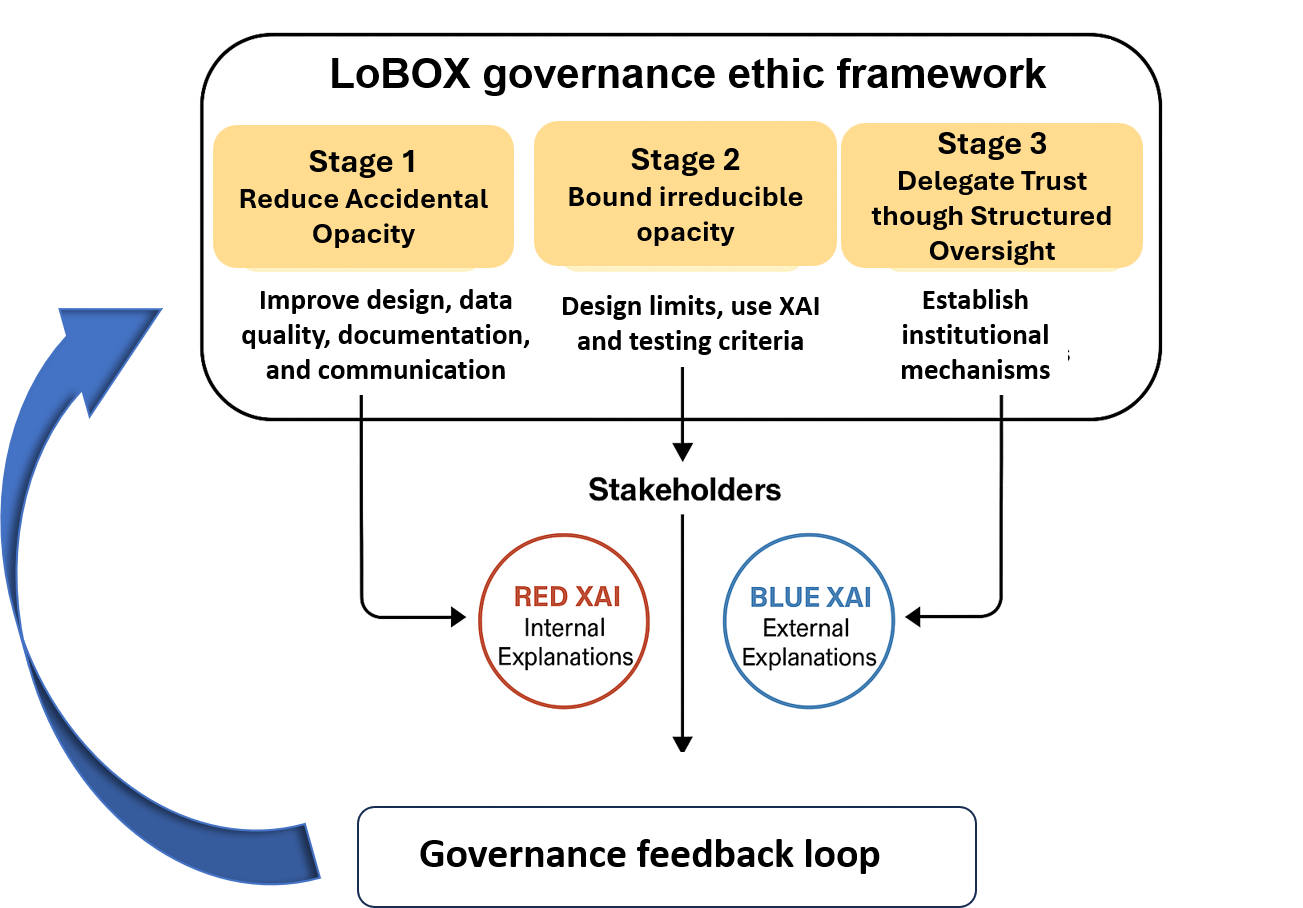}
    \caption{ Governance feedback loop: Reinforcing adaptive trust in LoBOX}
    \label{fig:LoBOX}
\end{figure*}

As shown in Figure \ref{fig:LoBOX}, this loop connects the three core stages of the LoBOX framework, reduction, bounding, and delegation, with recursive oversight and stakeholder participation, illustrating how ethical opacity management can evolve without compromising legitimacy.

\subsection{Beyond Utilitarian Trust: Institutionalizing the Leap of Faith}
\label{sec:LoBOX-2}

The normative foundation of LoBOX is directly derived from the critique of utilitarian trust \cite{van2023critique}: the assumption that trust is formed through rational assessments of risk and utility. Although such a model may be applicable in isolated or one-to-one interactions, it fails under the complexity of sociotechnical systems like AI, where asymmetries of information, unpredictability, and dispersed responsibility undermine calculative trust.
Their empirical work identifies three systemic inconsistencies that challenge utilitarian trust.

\begin{enumerate}
    \item Lack of sufficient information to assess risks or returns
      \item  Unstable and unpredictable behaviour among stakeholders
        \item Blurred lines of responsibility and authority
\end{enumerate}

In such environments, trust cannot be grounded in individual rationality. Instead, it requires what they call a “leap of faith”—a decision to trust without full comprehension. Crucially, they argue that this leap must not be a private or impulsive act; it must be ethically scaffolded by institutional design.

This is precisely where LoBOX intervenes. Rather than treating the leap of faith as a vulnerability, it reframes it as a governance task:

\begin{itemize}

    \item  Stage 1 reduces the need for unwarranted trust by eliminating avoidable opacity through clearer interfaces, documentation, and data quality.
    \item  Stage 2 contains irreducible opacity with role-sensitive explanation strategies.
    \item  Stage 3 shifts the source of trust from comprehension to institutional credibility, supported by audits, redress mechanisms, inspection and AI safety models, and ethical oversight.

\end{itemize}

In doing so, LoBOX operationalizes Van Rietschoten and Van Bommel’s \cite{van2023critique} call to move beyond the calculative trust models. It redistributes epistemic and ethical responsibility through governance structures that enable trust even in the absence of full system transparency. By embedding role-based accountability into opaque environments, LoBOX provides a pragmatic and principled foundation for responsible AI deployment, for responsible AI systems design \cite{herrera-poyatos2024}.

\subsection{Core Design Principles: What Makes the LoBOX Framework Different}
\label{sec:LoBOX-3}

LoBOX is built on four foundational features that enable ethical oversight under conditions of opacity. Each reflects a departure from traditional transparency-driven models, emphasizing a more pragmatic, relational, and institutionally grounded approach to explainability and trust in sociotechnical systems \cite{ananny2018seeing,selbst2019fairness,freiman2025opacity}.

\begin{itemize}

 \item 	\textit{Role-sensitive Explainability:} Different stakeholders, developers, regulators users, receive explanations aligned to their needs and capacities. LoBOX recognizes that AI systems interact with diverse stakeholders whose needs and interpretive capacities vary widely. Developers and engineers require technical insights into model internals for debugging (e.g., feature attribution, training diagnostics), whereas end-users, such as patients or applicants, benefit more from high-level context-sensitive justifications that promote understanding and contestability \cite{arrieta2020explainable}. Regulators, on the other hand, need oversight mechanisms that enable systemic evaluation and risk assessment. To meet these differentiated demands, LoBOX operationalizes the RED/BLUE XAI model \cite{biecek2024explain}, which distinguishes between internal technical explanation and external moral intelligibility. This ensures that the explanation is not universalized but is tailored to the role of the stakeholder, the epistemic authority, and the ethical proximity.

 \item 	\textit{Stakeholder-Calibrated Design:}  Governance intensity scales with risk and user vulnerability. LoBOX adopts a risk-sensitive design philosophy in which the intensity of governance increases with the system criticality and the vulnerability of the user. Inspired by normative models of proportionality \cite{raji2020closing,fraser2022ai} , this feature ensures that high-risk systems, such as those used in criminal justice or social benefits allocation, are subject to more rigorous oversight than lower-stakes applications, such as product recommendation. This calibration also acknowledges power asymmetries and institutional stakes, prioritizing ethical scrutiny where the consequences of failure are greatest \cite{selbst2019fairness}. Move governance from static checklists to context-aware intervention, reinforcing the need for adaptable regulatory architectures.

 \item \textit{Trustworthy opacity:} 	 Opacity is managed rather than eradicated; Justified opacity replaces unrealistic transparency. Rather than defining opacity as a design flaw, LoBOX emphasizes that some opacity is both inevitable and ethically permissible, particularly when rooted in technical complexity or privacy-preserving constraints \cite{lipton2018mythos,boge2022two}. The key is whether the opacity is justified and governable. LoBOX promotes "trustworthy opacity" by embedding oversight through explainability tools (e.g., SHAP, LIME), deployment boundaries, and role-specific accountability structures. Opacity, in this model, becomes ethically acceptable when it is documented, monitored and contestable, an idea that aligns with contemporary XAI critiques and legal analogs in systems such as courts or public health \cite{fraser2022ai, de2024black}.
 
 \item 	\textit{Governed Leap of Faith: }  The “leap of faith” becomes an institutionally scaffolded act, not a blind or passive trust. Drawing on the critique of utilitarian trust \cite{van2023critique} , LoBOX reframes the “leap of faith” not as a psychological deficiency, but as an inevitable governance condition in complex AI ecosystems under the name of “leap of belief”. When users lack the cognitive or informational means to directly evaluate AI systems, trust must be institutionally built through mechanisms such as audits, recourse channels, and public accountability boards \cite{flores1998creating,mollering2001nature}. In this way, LoBOX turns opacity from a source of mistrust into a structured opportunity to build legitimacy. 

\end{itemize}
\subsection{Role-Sensitive Explainability Through RED/BLUE XAI}
\label{sec:LoBOX-4}

A defining strength of the LoBOX governance ethic framework is its integration with the RED/BLUE XAI model proposed by Biecek and Samek (2024), which offers a structured approach to aligning explanation strategies with audience-specific needs. This integration allows LoBOX to operationalize role-sensitive explainability throughout the AI lifecycle, ensuring that each stakeholder receives the type and level of information necessary for ethical scrutiny, decision making, or redress.

\begin{itemize}

 \item 	\textit{RED (Research, Explore, Debug) XAI} provides technical transparency targeted at internal actors such as developers, data scientists, compliance officers, and institutional auditors. These explanations include performance diagnostics, input-output attribution (e.g., SHAP or LIME), error propagation, or counterfactual examples. The purpose is to support debugging, validation, and internal auditability, helping technical teams manage both accuracy and accountability under complex model conditions. RED XAI is most effective during the development and training phases, where access to detailed internals of the model supports oversight and refinement. RED XAI promotes technical auditability.
 
 \item 	\textit{	BLUE (responsiBle, Legal, TrUst, Ethics)} XAI addresses external stakeholders, including end-users, affected citizens, patients, regulators, or civil society advocates, by offering morally intelligible and practically actionable explanations. These may take the form of natural language rationales, impact summaries, appeal pathways, or simplified outcome justifications. The goal is not to make the model transparent in a technical sense, but to make its decisions comprehensible and challengeable in socially meaningful terms. BLUE XAI supports procedural fairness, contestability, and inclusive understanding in public-facing systems. BLUE XAI ensures public accountability.

 \end{itemize}
 
 By distinguishing between what needs to be explained, to whom, for what purpose, and at what level of abstraction, LoBOX avoids the common pitfall of “one-size-fits-all” explainability (Lipton, 2018; Arrieta et al., 2020). It ensures that the explanation is not only technically rigorous, but relationally adequate, and responsive to the epistemic roles and moral stakes of diverse actors.

Moreover, the RED/BLUE XAI integration enables LoBOX to satisfy both compliance demands and public legitimacy requirements. RED XAI facilitates traceability and documentation for legal or institutional review (e.g., Article 9 of the EU AI Act on risk management), while BLUE XAI aligns with human-centric mandates for transparency and contestability (e.g., Article 13 on user comprehension or Article 29 on rights impacts).

In sum, RED XAI serves internal stakeholders by supporting debugging and validation, while BLUE XAI ensures that external stakeholders receive morally intelligible and contestable explanations.

\subsection{Translating Ethics into Regulation: LoBOX Framework in Legal Context}
\label{sec:LoBOX-5}

The EU AI Act introduces a tiered regulatory framework that calibrates obligations based on the level of system risk. High-risk systems, in particular, are required to maintain an auditable risk management system (Art. 9, 11), provide clear and understandable information to users (Art. 13; Art. 52 1 for GPAI), assess and mitigate the impacts of fundamental rights (Art. 29), and ensure human oversight and post-market monitoring (Art. 14, 17, 61).
The LoBOX governance ethic framework complements and reinforces these requirements by offering a structured pathway for ethical oversight that aligns with the normative goals of the AI Act. When viewed through two complementary lenses: risk-tiered governance and role-sensitive explainability, the LoBOX helps translate legal mandates into actionable system design and oversight strategies.

\begin{itemize}

 \item 	\textit{Stage 1 (Reduce Accidental Opacity)}  directly supports Articles 9 and 11, by encouraging early-stage risk documentation, usability testing, and transparency-by-design practices.
 
 \item 	\textit{Stage 2 (Monitor Per Se Opacity)} aligns with Articles 13 and 29, by operationalizing stakeholder-appropriate explanation tools (e.g., RED/BLUE XAI) and bounding opacity within ethical use constraints.
 \item 	\textit{Stage 3 (Delegate Trust via Institutional Oversight)} is especially relevant to Articles 14 and 61, which emphasize human oversight, contestability, and the ability to override or monitor system behavior after deployment.
  
  \end{itemize}

In this way, LoBOX offers a governance-by-design alternative to purely technical or transparency-centric approaches. Rather than assuming that explanation alone guarantees compliance or legitimacy, the framework embeds structured justification, stakeholder calibration, and institutional accountability into the deployment of AI systems. It provides a robust ethical framework that meets both the letter and the spirit of emerging AI regulation in Europe.

\section{Application scenarios }
\label{sec:App}

This section presents three real-world scenarios that demonstrate how the LoBOX governance ethic framework can be applied across diverse AI deployment contexts. Each example illustrates different challenges in managing opacity - from clinical risk, systemic bias in financial systems, to fairness in educational access - and highlights how LoBOX's three-stage framework addresses these challenges through role-sensitive explainability, institutional oversight, and, when necessary, ethical system redesign. Table \ref{tab:App} summarizes the focus and governance implications of each case.

\begin{table}[th!]
\centering
\caption{ . Summary of LoBOX applications}
\label{tab:App}
\begin{tabular}{|p{2.4cm}|p{3cm}|p{3.4cm}|p{3.4cm}|}
\hline

Example & Opacity Challenge & LoBOX Focus Areas & Outcome \\ \hline

High-risk Clinical Decision System & Per se opacity in triage model; risks to patient safety & All stages, with emphasis on Stage 2 for explainability checkpoints and Stage 3 for ethics board review & Preserved clinical judgment, embedded human review, strengthened trust \\ \hline

Automated Loan Approval at Scale & Unexplainable denials, systemic bias, limited customer comprehension & Stage 3 essential due to limits of explainability; triggered systemic audits and fairness reviews & Reframed trust through institutional mechanisms, enhanced procedural fairness \\ \hline

Automated Loan Approval at Scale & Opaque essay scoring, fairness concerns, demographic disparities  & Stage 3 leads to model redesign, data diversification, and explainability-by-design interface & Redesigned system architecture to improve intelligibility and inclusivity \\ \hline

\end{tabular}
\end{table}

\subsection{Case 1: LoBOX Framework in a High-Risk Clinical Decision System}
\label{sec:App-1}

To demonstrate the practical application of the LoBOX framework, consider the case of a national healthcare agency that is deploying an AI system to prioritize access to specialist medical treatment. The system is trained in historical referral data and patient outcomes and is designed to recommend priority levels for new cases. Given the context of high-risk scenarios, explainability and trust are critical.

In Stage 1 – Reduce Accidental Opacity, the agency implements a user-centered interface for medical personnel, including tools explaining output categories, natural language descriptions of recommendations, and accessible documentation for how priority levels are determined. RED XAI techniques are used internally for debugging and auditing, ensuring that the behavior of the system is in line with the intended policy goals.

In Stage 2 – Bound Per Se Opacity, the agency uses SHAP and LIME to generate interpretability checkpoints for decision justification, even though the underlying model is a complex ensemble. These tools are applied during model validation and periodically after deployment. The agency defines strict limits for the use of the model: limiting it to non-emergency referrals and mandating human review for outlier predictions. Ethical restrictions are formalized in the documentation and reviewed by a clinical ethics board.

In Stage 3 – Delegate Trust through Institutional Oversight, patients and their advocates are provided with BLUE XAI explanations for their triage outcomes. These include a summary of the factors that influence and appeal instructions. The agency also establishes an oversight committee that includes clinicians, legal experts, and patient representatives. The committee conducts regular audits, reviews edge cases, and is empowered to recommend model suspension if fairness concerns arise.

Through this staged governance approach, LoBOX enables the agency to manage opacity not by forcing full transparency, but by making the system accountable, understandable to each stakeholder group, and embedded in credible institutional safeguards. It shows how LoBOX can be used not only as an ethical guide but also as a compliance tool aligned with emerging regulatory demands such as the EU AI Act.

For example, suppose that the model flags a 62-year-old patient with chronic kidney disease and intermittent hypertension as 'low priority' for nephrology referral. Clinically, this recommendation raises red flags due to the patient's age and comorbidities. SHAP analysis reveals that the model reduced kidney function due to confusion in training data, where younger patients with similar lab profiles often recovered without intervention. In this case, the system-generated output triggers an explainability checkpoint, weakening the case for mandatory human review. The attending physician overrides the model's suggestion and initiates a referral. This interaction is recorded and later reviewed by the ethics board as part of routine audit procedures. The case highlights how XAI tools, bounded autonomy, and clinical review work together to manage per se opacity in a way that preserves clinical judgment and supports ethical containment.

This example illustrates how tools, bounded autonomy, and clinical review work together to manage per se opacity in a way that preserves clinical judgment and supports ethical containment. Crucially, it underscores the importance of human–AI collaboration in high-risk environments. LoBOX not only structures technical oversight, but also reinforces the normative imperative that AI systems must support - not replace - human reasoning. By integrating RED/BLUE XAI, LoBOX ensures that explanations are context-sensitive and role-appropriate, allowing clinicians to understand, question, or override recommendations. This prevents overreliance on algorithmic outputs and re-centers professional discretion in critical decisions. In this light, explainability is not only a compliance tool, but a safeguard for ethical co-decision-making: anchoring institutional trust in shared accountability between humans and machines.

\subsection{Case 2: LoBOX Framework in Automated Loan Approval on Scale}
\label{sec:App-2}

Consider a national financial regulator overseeing a consortium of banks that implement a deep learning model to automate loan approval decisions across multiple demographics and regions. The system is trained on years of credit data, income records, and behavioral indicators. Although it performs well in aggregate, the complexity of the model, combined with proprietary components and retraining cycles, renders it largely inscrutable, even to internal developers.
In this context, RED XAI tools, such as SHAP or counterfactuals, are used to attempt post hoc justification for loan denials. However, inconsistencies emerge. Explanations for similar applicants vary widely and, even when technically accurate, they are incomprehensible to customers or yield little actionable feedback. Worse, marginalized applicants are disproportionately affected, raising concerns of systemic bias that is not traceable to a single variable or decision point. This shows a limit of explainability. Despite the use of XAI, the opacity is not significantly reducible at the individual level.

Here, stage 3 of the LoBOX becomes essential. The regulatory authority mandates that all participating banks adopt standardized oversight protocols, including:

\begin{itemize}
  
  \item  	Public reporting on approval disparities across protected categories.
  \item  	A cross-institutional ethics review board was empowered to investigate model behavior.
  \item  	An ombudsperson mechanism that allows applicants to contest decisions.
  \item  	Ongoing audits using synthetic data to probe bias and decision drift.
  \item  	Formal policy boundaries restricting model updates without recertification.
\end{itemize}

Rather than relying on individual users or even developers to “understand” the system, LoBOX reframes trust as an institutionally scaffolded process. Transparency is replaced by contestability, procedural fairness, and embedded human oversight. This example shows how LoBOX enables governance where technical explainability falls short, feedback,  ensuring that trust is not blind, but anchored in systemic accountability.

When stage 3 triggers reengineering. In systems where per se opacity cannot be meaningfully explained to stakeholders, even with RED/BLUE XAI, LoBOX does not stop at containment. Its third stage, delegate trust through institutional oversight, also serves as a trigger for the AI system redesign and improvement loop. When oversight mechanisms (e.g. audits, redress pathways, public scrutiny) consistently reveal unresolved fairness issues, explainability failures, or stakeholder dissatisfaction, these findings feed back into governance: Demand for retraining, architecture revision, or feature re-assessment.

In this sense, Stage 3 becomes both a containment mechanism and a redesign catalyst.

\subsection{Case 3: LoBOX Framework to Automate University Admissions Screening}
\label{sec:App-3}

A national university system adopts an AI tool to pre-screen undergraduate applicants for large-scale programs. The model analyzes GPA, test scores, recommendation letters (via NLP), and demographic data. While explainability tools are built in, applicants who appeal decisions often receive generic explanations (e.g., 'Your application score was below the threshold') and are given no insight into how qualitative materials (e.g., essays or letters) were processed.

Over time, the oversight bodies observe the following low quality data.

\begin{itemize}
  
  \item Underrepresentation of applicants from rural or low-income backgrounds,
 \item Frequent appeals clustered around specific high schools,
 \item Inconsistent correlations between human-reviewed decisions and automated outputs.
\end{itemize}

Despite attempts to use RED XAI, the reliance of the system on proprietary NLP embeddings and the lack of diversity of training data limit the usefulness of post hoc explanations. At this point, Stage 3 governance mechanisms initiate a full system review, leading to:

\begin{itemize}

\item Suspension of the NLP pipeline for essay scoring until interpretability improves,
 \item Inclusion of human reviewers in cases flagged as borderline by the model,
 \item Commissioning of new training data that better reflects applicant diversity,
 \item Design of an “explainability by design” admissions tool that outputs scoring rationales in plain language from the outset.
\end{itemize}

This scenario illustrates how the institutional feedback loops supported by LoBOX don’t just constrain opacity—they guide redesign toward systems that are more accountable, inclusive, and intelligible.

When Stage 3 reveals that even bounded opacity is unacceptable due to persistent bias, public resistance, or legal non-compliance, LoBOX calls for:

\begin{itemize}
 \item Retraining with new or diversified datasets,
 \item Algorithmic simplification or hybrid models,
 \item Increased stakeholder participation in feature selection and validation,
 \item New interface design that prioritizes contestability over automation.
\end{itemize}

Thus, LoBOX not only manages the existing opacity but also enables the evolution of an iterative, ethically grounded AI system.

\section{Global Adaptability and Governance and Contextual Trust}
\label{sec:society}

The effectiveness of LoBOX hinges on its adaptability in diverse governance landscapes. Cultural norms and institutional trust vary widely, influencing how opacity and accountability are perceived. This section examines how LoBOX can be tailored to reflect these contextual differences.

Although the LoBOX emphasizes institutional scaffolding and stakeholder-relative justification, its applicability must be sensitive to variations in cultural and institutional trust across geopolitical contexts. For example, in many regions of the Global South, trust in formal institutions can be limited due to historical inequities, weak regulatory enforcement, or perceived lack of transparency. In contrast, European governance models, such as the EU AI Act, presuppose relatively high baseline trust in legal and bureaucratic institutions. Adapting LoBOX across these diverse contexts requires not only translating its principles into different legal systems, but also accounting for sociopolitical realities, levels of digital literacy, and cultural understandings of authority and accountability. 

Future work should explore how trust calibration within LoBOX can incorporate localized norms, community oversight structures, and participatory mechanisms - ensuring that ethical opacity governance remains equitable and context-aware on a global scale. The RED/BLUE distinction further enhances the portability of LoBOX by aligning the explanation strategies with localized expectations of authority, expertise, and user empowerment.

We must mention that while LoBOX aligns closely with the EU AI Act and its risk-based regulatory architecture, future applications should account for the diversity of global AI governance regimes. For example, the U.S. National Institute of Standards and Technology (NIST) AI Risk Management Framework emphasizes voluntary, industry-led practices, while the OECD AI Principles foreground human-centered values, transparency, and international interoperability. Furthermore, the recent report by the United Nations Governing AI for Humanity (2024) proposed a multilateral vision centered on global cooperation, equity, and sustainable development in AI deployment. Adapting LoBOX across these regulatory landscapes will require translating its core elements, such as role-sensitive explanation, bounded opacity, governance elements, and institutional trust, into varied legal, cultural, and institutional contexts. Comparative analysis across such frameworks can improve LoBOX flexibility and affirm its value as a portable governance ethic for managing opacity in globally deployed AI systems.

In this context, we must point out that while this paper establishes LoBOX as a principled and scalable framework for opacity governance, future research should deepen its application across emerging domains. On the societal side, this includes operationalizing stakeholder participation through civic AI councils or democratic oversight mechanisms and evaluating how role-sensitive explainability impacts equity and justice, especially in marginalized communities. 

To ensure that LoBOX supports not only formal institutions, but also democratic inclusion, practical implementations should explore how participatory mechanisms - such as civic AI councils, citizen juries, or stakeholder panels - can be embedded across its governance stages. These structures could serve as feedback nodes within the governance loop, validating whether RED/BLUE explanation strategies are intelligible, fair, and locally legitimate. This participatory layer is particularly vital in contexts where trust in centralized oversight is low or contested. 

Moreover, LoBOX has potential as a justice enabling tool by tailoring explanations to marginalized communities and supporting recourse mechanisms. It can help reduce systemic bias and restore agency in AI-impacted decision making.

\subsection*{Limitations and Theoretical Scope}

This paper presents the LoBOX governance ethic framework as a conceptual proposal based on normative reasoning and policy alignment, focused on the per se opacity but not yet fully operationalized. As a theoretical model, LoBOX provides an ethical and structural lens for managing opacity in AI systems. However, translating this model into real-world governance infrastructures is not trivial. It requires a deep engagement with legal systems, institutional capabilities, stakeholder dynamics, and enforcement mechanisms. Future work must move beyond conceptual scaffolding to build practical toolkits, policy interfaces, and implementation protocols capable of embedding LoBOX within diverse regulatory, organizational, and technical contexts.

\subsection*{Toward Global AI Governance: Reflections}

Recent work emphasizes the critical need for robust international AI governance that addresses both opportunities and risks. Taeihagh (2021) \cite{taeihagh2021governance} outlines the fragmented and uneven global landscape of AI oversight, calling for adaptive governance models that can respond to emerging sociotechnical challenges. Goos and Savona (2024)  \cite{goos2024governance} similarly stress that AI governance must balance innovation and risk mitigation by embedding coordination, capacity-building, and ethical safeguards throughout the AI lifecycle. The United Nations (2024) \cite{UN2024} builds on this by advocating for inclusive global frameworks, highlighting representation gaps, interoperability challenges under title \textit{"Governing AI for Humanity. Final Report}, and the urgency of a coherent governance infrastructure. Collectively, these perspectives reinforce the urgency of extending LoBOX beyond a Western-centric policy context and embedding it within a globally networked governance system grounded in equity, transparency, and institutional credibility.

Building on these global visions, several recurring governance challenges emerge that are particularly salient for implementing frameworks like LoBOX. Taeihagh \cite{taeihagh2021governance} identifies coordination deficits and regulatory fragmentation as central barriers, especially when AI development and deployment cross national borders and sectoral boundaries. Goos and Savona \cite{goos2024governance} further highlight three persistent tensions in AI governance: (1) the difficulty of achieving coherence between national innovation strategies and global ethical imperatives; (2) capacity asymmetries between regulators and technology developers; and (3) temporal mismatches between rapid AI deployment and the slow pace of institutional adaptation. The United Nations Advisory Body \cite{UN2024} underscores the geopolitical and social equity stakes in AI governance, arguing for multilateral oversight mechanisms that ensure marginalized voices are represented in shaping AI futures. 

For LoBOX to serve as a globally adaptable governance ethic, it must be able to address these structural asymmetries and embedding mechanisms for cross-jurisdictional coordination, procedural legitimacy, and inclusive stakeholder engagement. In this light, LoBOX should not only be treated as a governance toolkit but also as a normative infrastructure - capable of supporting adaptive regulation, capacity building, and equitable participation in the governance of increasingly opaque and impactful AI systems.

\subsection*{Embedding Audience-Calibrated Explainability in Global Governance}

To realize the LoBOX vision discussed, future research must go beyond static visualizations and generic rationales. It must develop audience-calibrated explanations embedded in procedural accountability, governance and human-AI collaboration to increase human-capabilities to take decisions. The LoBOX represents a step toward this future, offering a roadmap to manage opacity through structured justification and ethical alignment.

In this context, explainability should be understood not only as a technical feature but as a normative bridge between AI system creators and communities affected by these technologies. It constitutes the mechanism through which automated decisions can be scrutinized, contextualized, and held accountable, thus enabling informed and equitable human-AI interaction. Lacking such interpretive pathways risks more than confusion; it threatens the erosion of institutional trust and the displacement of democratic oversight by opaque computational processes. Alam et al.~\cite{alam2024automation} argue that sustainable integration of AI in critical sectors demands an augmentation approach, one that centers human reasoning, professional accountability, and cooperative decision making over full autonomy.

Looking ahead, the promise of responsible AI does not rest on predictive accuracy alone but on the system's ability to be explainable, interrogable, and normatively governed. This paper advocates for a shift in both research and policy: developers must treat explainability as a foundational design principle, not an afterthought; policymakers must embed intelligibility into regulatory standards; and society at large must resist any default toward unaccountable automation.  The future of human-AI collaboration depends on governance architectures that prioritize transparency, contestability, and public legitimacy from the ground up, and on a collective societal refusal to normalize the absence of effective opacity governance. Ensuring that AI systems operate within ethically responsible boundaries requires ongoing vigilance, institutional responsibility, and meaningful stakeholder engagement at every level.

Future work should explore domain-specific LoBOX deployments through empirical case studies, practical regulatory co-design, and integration with automated audit pipelines. This includes evaluating RED/BLUE XAI strategies in real-time settings, refining oversight triggers, and embedding the LoBOX into sector-specific legal and ethical standards. This vision also aligns with Mollick’s (2024) call for 'cointelligence' \cite{mollick2024co}, where humans and AI systems collaborate as complementary agents, achieving results that neither could achieve alone.

\section{Conclusion: Technical Implications and Future Pathways }
\label{sec:con}

Building on the societal dimensions outlined in Section 5, this final section reflects on LoBOX’s technical implications and presents conclusions and future directions from a XAI technical perspective, and as complement of previous section.

"From transparency ideals to ethical governance" and "from transparency to governable opacity" capture the core vision of the LoBOX framework: a scalable approach for managing opacity in AI systems. Rather than treating transparency as an ethical goal in itself. LoBOX reframes opacity as a condition to be ethically governed through stakeholder-sensitive explanations, institutional oversight, and structured justification. It challenges utilitarian assumptions that trust is a function of rational comprehension and instead grounds legitimacy in institutional credibility, accountability mechanisms, and contestability procedures.

Looking ahead, explainability must improve human AI decision-making by enabling systems to act as cognitive partners, particularly in high-risk, high-uncertainty domains such as law, medicine, and finance. When responsibly designed, XAI enables humans not only to interpret AI outputs but also to challenge, contextualize, and act upon them. In this role, XAI helps systems perform not only as well as humans but often better, amplifying human judgment in ways that are transparent, contestable, and aligned with public values.

By shifting the focus from universal transparency to role-sensitive governance, LoBOX reframes the central challenge of trustworthy AI: not how to make everything visible, but how to ensure that what remains opaque is ethically controllable. In doing so, it contributes not only a technical toolkit, but also a governance ethic rooted in realism, responsibility, and procedural justice.

As AI systems become more complex and consequential, efforts to fully interpret their inner workings often fall short. LoBOX addresses this reality not by abandoning oversight, but by redistributing it: establishing trust in governance architectures that respond to risk, role, and moral responsibility. 

In an era of algorithmic opacity, restoring public confidence requires more than technical disclosure. As the limitations of individual understanding and utilitarian trust become evident, a new ethical foundation must emerge, one that balances institutional design, stakeholder roles, and regulatory mandates. 

As AI systems scale and evolve into general-purpose models, such as large language models or generative systems, the challenge of implementing LoBOX will intensify. Recursive auditability, while central to the framework, may require automation support, integration into MLOps pipelines, and real-time monitoring capabilities. Exploring how LoBOX principles can be embedded in the AI lifecycle, across model design, deployment, and deprecation, offers a promising direction for future work. Additionally, generative AI introduces new complexities in explainability, as the model output is stochastic and content-based. Adapting role-sensitive explanations and bounding mechanisms to these systems could form the basis of a future extension of LoBOX tailored to generative architectures and AI-human co-intelligence.


\section*{Acknowledgments}
This publication is part of the project “Ethical, Responsible,
and General Purpose Artificial Intelligence: Applications In
Risk Scenarios” (IAFER) Exp.:TSI-100927-2023-1 funded
through the creation of university-industry research programs
(Enia Programs), aimed at the research and development of artificial intelligence, for its dissemination and education within
the framework of the Recovery, Transformation and Resilience
Plan from the European Union Next Generation EU through
the Ministry of Digital Transformation and the Civil Service.
 This work was also partially supported by Knowledge Generation Projects, funded by the Spanish Ministry of Science, Innovation, and Universities of Spain under the project PID2023-150070NB-I00.

\section*{Declaration of AI-assisted technologies in the writing process}

During the preparation of this work, the author used large-language models to improve the readability and language of the manuscript. After using this tool/service, the author reviewed and edited the content as needed and assumed full responsibility for the content of the published article.

\bibliographystyle{apacite}

\bibliography{bibliography}

\end{document}